\documentclass[aps,preprint]{revtex4}%
\usepackage{amsfonts}
\usepackage{amsmath}
\usepackage{amssymb}
\usepackage{graphicx}%
\providecommand{\U}[1]{\protect\rule{.1in}{.1in}}
\newtheorem{theorem}{Theorem}
\newtheorem{acknowledgement}[theorem]{Acknowledgement}

\begin{document}
\title{Generalized Galilean Algebras and Newtonian Gravity}
\author{N. Gonz\'{a}lez$^{1}\thanks{nicolas.gonzalez@physik.uni-muenchen.de}$, G.
Rubio$^2\thanks{gurubio@udec.cl}$, P. Salgado$^2\thanks{pasalgad@udec.cl}$, S.
Salgado$^2\thanks{sesalgado@udec.cl}$}
\affiliation{$^{2}$Departamento of F\'{\i}sica, Universidad de Concepci\'{o}n, Casilla
160-C, Concepci\'{o}n, Chile}
\affiliation{$^{1}$Arnold Sommerfeld Center for Theoretical Physics,
Ludwig-Maximilians-Universit\"{a}t M\"{u}nchen, Theresienstra\ss e 37, D-80333
Munich, Germany}

\begin{abstract}
The non-relativistic versions of the generalized Poincar\'{e} algebras and
generalized $AdS$-Lorentz algebras are obtained. This non-relativistic
algebras are called, generalized Galilean algebras type I and type II and
denoted by $\mathcal{G}\mathfrak{B}_{n}$ and $\mathcal{G}\mathfrak{L}_{_{n}}$
respectively. Using a generalized In\"{o}n\"{u}--Wigner contraction procedure
\ we find that the generalized Galilean algebras type I can be obtained from
the generalized Galilean algebras type II. The $S$-expansion procedure allows
us to find the $\mathcal{G}\mathfrak{B}_{_{5}}$ algebra from the Newton Hooke
algebra with central extension. The procedure developed in Ref. \cite{newton}
allow us to show that\ the nonrelativistic limit of the five dimensional
Einstein--Chern--Simons gravity is given by a modified version of the Poisson
equation. The modification could be compatible with the effects of Dark
Matter, which leads us to think that Dark Matter can be interpreted as a
non-relativistic limit of Dark Energy.

\end{abstract}
\maketitle

\section{\textbf{Introduction}}

In Refs. \cite{gr-chs, gr-bi} was shown that the $S$-expansion procedure
allows to construct Chern--Simons gravities in odd dimensions invariant under
an algebra referred as $\mathfrak{B}_{m}$ algebra and Born--Infeld gravities
in even dimensions \cite{tron, ban, ban1,zan} invariant under a certain
subalgebra of the $\mathfrak{B}_{m}$ algebra,\ leading to general relativity
in a certain limit. The $\mathfrak{B}_{m}$ algebras, which could be also
called `generalized Poincar\'{e} algebras', was constructed from $AdS$-algebra
and a particular semigroup denoted by $S_{E}^{(N)}=\left\{  \lambda_{\alpha
}\right\}  _{\alpha=0}^{N+1}$, which is endowed with the multiplication rule
\ $\lambda_{\alpha}\lambda_{\beta}=\lambda_{\alpha+\beta}$ if \ $\alpha
+\beta\leq N+1;$ $\lambda_{\alpha}\lambda_{\beta}=\lambda_{N+1}$ if
\ $\alpha+\beta>N+1$.

In Ref. \cite{ssep} was shown that the so-called $AdS$-Lorentz algebra
$\mathfrak{so}\left(  D-1,1\right)  \oplus\mathfrak{so}\left(  D-1,2\right)  $
algebra \cite{soro1, soro2, soro3} in $D$ dimensions can be obtained from
$AdS$-algebra $\mathfrak{so}\left(  D-1,2\right)  $ by means of the
$S$-expansion procedure with a semigroup which is kown as $S_{\mathcal{M}%
}^{(2)}$.\ This $AdS$-Lorentz algebra is related to the so called Maxwell
algebra \cite{bacry, schr} via a contraction process \cite{luk}. \ 

Recently was shown in Ref. \cite{salg1} that the resonant $S$-expansion of the
$AdS$ Lie algebra leads to a generalization of the $AdS$-Lorentz algebra when
it is used $S_{\mathcal{M}}^{(N)}=\left\{  \lambda_{\alpha}\right\}
_{\alpha=0}^{N}$ as semigroup, which is endowed with the multiplication
rule\ $\lambda_{\alpha}\lambda_{\beta}=\lambda_{\alpha+\beta}$ if
\ $\alpha+\beta\leq N;$ $\lambda_{\alpha}\lambda_{\beta}=\lambda_{\alpha
+\beta-2\left[  (N+1)/2\right]  }$ if \ $\alpha+\beta>N$. \ These algebras are
called generalized $AdS$-Lorentz algebras. \ In this same Ref. \cite{salg1}
was found that a generalized In\"{o}n\"{u}--Wigner contraction of the
generalized $AdS$-Lorentz algebras provides the so called generalized
Poincar\'{e} algebras,$\ \mathfrak{B}_{m}$.

On the other hand, in Ref. \cite{newton} was shown how the Newton--Cartan
formulation of Newtonian gravity can be obtained from gauging the Bargmann
algebra, i.e. the centrally extended Galilean algebra.

This paper is organized as follows: In Section II it is shown that, using an
analogous procedure to that used in Ref. \cite{chern3} it is possible to
obtain the non-relativistic versions of the generalized Poincar\'{e} algebras
and generalized $AdS$-Lorentz algebras. The nonrelativistic algebras will be
called, \ generalized Galilean algebras type I and type II and denoted by
$\mathcal{G}\mathfrak{B}_{_{n}}$ and $\mathcal{G}\mathfrak{L}_{_{n}}$
respectively. In Section III it is shown that \ the generalized Galilean
algebras type I can be obtained by a generalized In\"{o}n\"{u}--Wigner
contraction of generalized Galilean algebras type II. \ In this section it is
also shown that the procedure of $S$-expansion allows us to find the
$\mathcal{G}\mathfrak{B}_{_{5}}$ algebra from the Newton Hooke algebra with
central extension. In Section IV it is show that the non-relativistic limit of
Einstein--Chern--Simons gravity is given by a modified version of the Poisson
equation. In Section V it is found that,\ using an analogous procedure to that
used in Ref. \cite{newton}, it is possible to find a generalization of the
Newtonian gravity. Finally our conclusions are presented in Section VI.

\section{\textbf{Generalized Galilean type }I\textbf{ }$(\mathcal{G}%
\mathfrak{B}_{_{n}})$\textbf{ and type }II\textbf{ }$(\mathcal{G}%
\mathfrak{L}_{_{n}})$ \textbf{algebras}}

The use of the procedure developed in Ref. \cite{chern3}, allow us to show
that it is possible to obtain the non-relativistic versions of the generalized
Poincar\'{e} algebras and of the generalized $AdS$-Lorentz algebras. The
nonrelativistic algebras will be called, generalized Galilean type I and type
II algebras and denoted by $\mathcal{G}\mathfrak{B}_{_{n}}$ and $\mathcal{G}%
\mathfrak{L}_{_{n}}$ respectively. We consider the particular cases $n=4,5$.

Consider now the non relativistic versions of the Maxwell and $\mathfrak{B}%
_{_{5}}$ algebras\textbf{. }Separating the spatial temporal components in the
generators $\left\{  P_{a},J_{ab},Z_{a},Z_{ab}\right\}  $, performing the
rescaling $K_{i}\longrightarrow c^{-1}J_{i0}$, $P_{i}\longrightarrow
R^{-1}P_{i}$, $H\longrightarrow cR^{-1}P_{0}-c^{2}M$, $Z_{i0}\longrightarrow
c^{-1}Z_{i0}$, $Z_{i}\longrightarrow R^{-1}Z_{i}$, $Z_{0}\longrightarrow
cR^{-1}Z_{0}-c^{2}N$ and then taking the limit $c$, $R\rightarrow\infty$, we
find that:

$(i)$ the generators of the non-relativistic version of the Maxwell algebra,
which we will denote by $\mathcal{G}\mathfrak{B}_{_{4}}$, satisfy the
following commutation relations%
\begin{align}
\lbrack J_{ij},J_{kl}] &  =\delta_{kj}J_{il}+\delta_{lj}J_{ki}-\delta
_{ki}J_{jl}-\delta_{li}J_{kj}\text{,}\nonumber\\
\lbrack J_{ij},K_{k}] &  =\delta_{jk}K_{i}-\delta_{ik}K_{j}\text{, \ \ }%
[K_{i},P_{j}]=-\delta_{ij}M\text{,}\nonumber\\
\lbrack J_{ij},P_{k}] &  =\delta_{jk}P_{i}-\delta_{ik}P_{j}\text{, \ }%
[K_{i},H]=-P_{i}\text{,}\nonumber\\
\lbrack J_{ij},Z_{kl}] &  =\delta_{kj}Z_{il}+\delta_{lj}Z_{ki}-\delta
_{ki}Z_{jl}-\delta_{li}Z_{kj}\text{,}\nonumber\\
\lbrack J_{ij},Z_{k0}] &  =\delta_{jk}Z_{i0}-\delta_{ik}Z_{j0}\text{,
\ \ }[P_{i},H]=\nu^{2}Z_{i0}\text{,}\nonumber\\
\lbrack Z_{ij},K_{k}] &  =\delta_{jk}Z_{i0}-\delta_{ik}Z_{j0}\text{.}%
\label{gb4}%
\end{align}
and $(ii)$ the generators of the non-relativistic version of the
$\mathfrak{B}_{_{5}}$ algebra \cite{gr-chs}, \cite{salg1} which we will denote
by $\mathcal{G}\mathfrak{B}_{5}$, satisfy the commutation relations%

\begin{align}
\lbrack J_{ij},J_{kl}]  &  =\eta_{kj}J_{il}+\eta_{lj}J_{ki}-\eta_{ki}%
J_{jl}-\eta_{li}J_{kj}\text{,}\nonumber\\
\lbrack J_{ij},K_{k}]  &  =\eta_{jk}K_{i}-\eta_{ik}K_{j}\text{, \ \ }%
[K_{i},P_{j}]=-\delta_{ij}M\text{,}\nonumber\\
\lbrack J_{ij},P_{k}]  &  =\eta_{jk}P_{i}-\eta_{ik}P_{j}\text{, \ }%
[K_{i},H]=-P_{i}\text{,}\nonumber\\
\lbrack J_{ij},Z_{kl}]  &  =\eta_{kj}Z_{il}+\eta_{lj}Z_{ki}-\eta_{ki}%
Z_{jl}-\eta_{li}Z_{kj}\text{,}\nonumber\\
\lbrack J_{ij},Z_{k0}]  &  =\eta_{jk}Z_{i0}-\eta_{ik}Z_{j0}\text{, \ \ }%
[K_{i},Z_{j}]=-\delta_{ij}N\text{,}\nonumber\\
\lbrack Z_{ij},K_{k}]  &  =\eta_{jk}Z_{i0}-\eta_{ik}Z_{j0}\text{, \ }%
[K_{i},Z_{0}]=-Z_{i}\text{,}\nonumber\\
\lbrack J_{ij},Z_{k}]  &  =\eta_{jk}Z_{i}-\eta_{ik}Z_{j}\text{, \ \ }%
[Z_{i0},P_{j}]=-\delta_{ij}N\text{,}\nonumber\\
\lbrack Z_{ij},P_{k}]  &  =\eta_{jk}Z_{i}-\eta_{ik}Z_{j}\text{, \ }%
[Z_{i0},H]=-Z_{i}\text{,}\nonumber\\
\lbrack P_{i},H]  &  =\text{ }\nu^{2}Z_{i0}\text{.} \label{gb5}%
\end{align}
where $\nu=c/R$ is a finite constant with$\ c$ the speed of light and $R$ the
universe radius. Following the same procedure used previously, it is possible
to find non-relativistic versions of the generalized AdS-Lorentz algebras,
which will be called generalized Galilean type II algebras and denoted and
$\mathcal{G}\mathfrak{L}_{_{n}}$. It is direct to show that \ using a
In\"{o}n\"{u}--Wigner contraction procedure we can obtain\ the $\mathcal{G}%
\mathfrak{B}_{_{n}}$ from $\mathcal{G}\mathfrak{L}_{_{n}}$.

\section{\textbf{ }$\mathcal{G}\mathfrak{B}_{_{n}}$\textbf{ algebras}}

\subsection{$\mathcal{G}\mathfrak{B}_{_{4}}$\textbf{ algebra from the
}$\mathcal{G}\mathfrak{L}_{_{4}}$\textbf{ algebra}}

From Ref. \cite{salg1} we know that the generalized Poincar\'{e} algebras can
be obtained from the generalized $AdS$-Lorentz algebras by means of a
generalized In\"{o}n\"{u}--Wigner contraction. This property of these
relativistic algebras is inherited by their corresponding nonrelativistic
algebras. We consider, as an example, the contraction of the $\mathcal{G}%
\mathfrak{L}_{_{4}}$ algebra. In fact, performing the following rescaling
$P_{i}\longrightarrow\lambda P_{i}$, $H\longrightarrow\lambda H$,
$M\longrightarrow\lambda M$, $Z_{i0}\longrightarrow\lambda^{2}Z_{i0}$,
$Z_{ij}\longrightarrow\lambda^{2}Z_{ij}$ \ of the $\mathcal{G}\mathfrak{L}%
_{_{4}}$ algebra provides in the limit $\lambda\longrightarrow0$ the
$\mathcal{G}\mathfrak{B}_{_{4}}$ algebra (\ref{gb4}). Similarly we can get the
$\mathcal{G}\mathfrak{B}_{_{5}}$ algebra from the $\mathcal{G}\mathfrak{L}%
_{_{5}}$ algebra.

\subsection{$\mathcal{G}\mathfrak{B}_{_{5}}$ \textbf{ algebra from the Newton
Hooke algebra}}

In Refs. \cite{gr-chs, gr-bi} was shown that the $S$-expansion procedure
allows to obtain the `generalized Poincar\'{e} algebras' $\mathfrak{B}_{n}$
from $AdS$-algebra. In this subsection it is shown that the procedure of
$S$-expansion allows us to find the $\mathcal{G}\mathfrak{B}_{_{5}}$ algebra
from the Newton Hooke algebra with central extension. A representation of
$AdS$ algebra is given by the matrices%
\[
\Gamma_{\mu\nu}=\frac{1}{4}\left[  \gamma_{\mu},\gamma_{\nu}\right]  \text{,}%
\]
where the matrices\textbf{ }$\gamma_{\mu}$\textbf{ }satisfy the Clifford
algebra\textbf{ }$\gamma_{\mu}\gamma_{\nu}+\gamma_{\nu}\gamma_{\mu}=2\eta
_{\mu\nu},$\textbf{ }with\textbf{ }$\mu,\nu:0,1,\cdot\cdot\cdot,5$\textbf{
}and where $\eta_{\mu\nu}=\mathrm{diag}\left(  -c^{2},1,1,1,1,-R^{2}\right)
$. The identification $J_{ij}=\Gamma_{ij},$ $\Gamma_{i0}=cK_{i},$ $\Gamma
_{i5}=RP_{i},$ $\Gamma_{05}=RP_{0}=Rc^{-1}\left(  H+c^{2}M\right)  $ and
$\Gamma_{\ast}=2M$ with $\Gamma_{\ast}=\gamma_{0}\gamma_{1}\gamma_{2}%
\gamma_{3}\gamma_{4}\gamma_{5}$, it leads to the commutation relations of the
Newton Hooke algebra with central extension Ref. \cite{nh}.

From Ref. \cite{iza} we find that the non-vanishing components of the
invariant tensor for $\mathfrak{so}(4,2)$ are given by%
\begin{align*}
\langle\Gamma_{\ast}\{\Gamma_{\mu\nu}\Gamma_{\rho\sigma}\Gamma_{\delta\tau
}\}\rangle &  =-8(Rc)^{2}\epsilon_{\mu\nu\rho\sigma\delta\tau}\text{,}\\
\langle\Gamma_{\ast}\{\Gamma_{\mu\nu}\Gamma_{\rho\sigma}\Gamma_{\tau
5}\}\rangle &  =-8(Rc)^{2}\epsilon_{\mu\nu\rho\sigma\tau5}\text{.}%
\end{align*}

Following an analogous procedure to that used in Ref. \cite{chern3}, we find
that the only nonzero components of the invariant tensor for the
$5$-dimensional Newton Hooke algebra with central extension.
\begin{align*}
\langle J_{ij}J_{kl}M\rangle &  =-16\epsilon_{ijkl}\text{,}\\
\langle J_{ij}P_{k}K_{l}\rangle &  =-16\epsilon_{ijkl}\text{.}%
\end{align*}

Following the definitions of Ref. \cite{salg2} (see also \cite{azcarr}) let us
consider the $S$-expansion of Newton Hooke algebra with central extension
using as semigroup $S_{E}^{\left(  3\right)  }=\left\{  \lambda_{0}%
,\lambda_{1},\lambda_{2},\lambda_{3},\lambda_{3}\right\}  $ endowed with the
multiplication rule \ $\lambda_{\alpha}\lambda_{\beta}=\lambda_{\alpha+\beta}$
if \ $\alpha+\beta\leq4;$ $\lambda_{\alpha}\lambda_{\beta}=\lambda_{4}$ if
\ $\alpha+\beta>4.$ After extracting a resonant and reduced subalgebra, one
finds the $\mathcal{G}\mathfrak{B}_{_{5}}$ algebra, given by (\ref{gb5}).
\ The invariant tensors for $\mathcal{G}\mathfrak{B}_{5}$ can be obtained from
Newton Hooke algebra with central extension. Using $VII$.2 from Ref.
\cite{salg2} we find%
\begin{align}
\langle J_{ij}J_{kl}M\rangle &  =-\frac{4}{3}\alpha_{1}lv\epsilon
_{ijkl}\text{, \ \ \ }\langle J_{ij}P_{k}K_{l}\rangle=-\frac{4}{3}\alpha
_{1}lv\epsilon_{ijkl}\text{, \ \ \ }\langle J_{ij}Z_{kl}M\rangle=-\frac{4}%
{3}\alpha_{3}lv\epsilon_{ijkl}\text{,}\nonumber\\
\langle Z_{ij}P_{k}K_{l}\rangle &  =-\frac{4}{3}\alpha_{3}lv\epsilon
_{ijkl}\text{, \ \ \ }\langle J_{ij}P_{k}Z_{l0}\rangle=-\frac{4}{3}\alpha
_{3}lv\epsilon_{ijkl}\text{, \ \ \ }\langle J_{ij}J_{kl}N\rangle=-\frac{4}%
{3}\alpha_{3}lv\epsilon_{ijkl}\text{,}\nonumber\\
\langle J_{ij}Z_{k}K_{l}\rangle &  =-\frac{4}{3}\alpha_{3}lv\epsilon
_{ijkl}\text{,} \label{tensor}%
\end{align}
where the constants $\alpha_{1}$ and $\alpha_{3}$ are dimensionless and the
factors $l$,$v$ are introduced to display the dimension of $\left\langle
\cdots\right\rangle $, where $l$ and $v$ are parameters of dimension length
and velocity respectively.

\section{\textbf{Non-relativistic limit of Einstein--Chern--Simons gravity}}

The five dimensional Chern--Simons lagrangian for the $\mathfrak{B}_{5}$
algebra is given by \cite{gr-chs}%
\begin{equation}
L_{\mathrm{ChS}}^{(5)}=\alpha_{1}l^{2}\varepsilon_{abcde}R^{ab}R^{cd}%
e^{e}+\alpha_{3}\varepsilon_{abcde}\left(  \frac{2}{3}R^{ab}e^{c}e^{d}%
e^{e}+2l^{2}k^{ab}R^{cd}T^{\text{ }e}+l^{2}R^{ab}R^{cd}h^{e}\right)  \text{,}
\label{1}%
\end{equation}
where $\alpha_{1}$, $\alpha_{3}$ are parameters of the theory, $l$ is a
coupling constant, $R^{ab}=d\omega^{ab}+\omega_{\text{ \ }c}^{a}\omega^{cb}$
corresponds to the curvature $2$-form in the first-order formalism. In Ref.
\cite{gomez} was considered that in the presence of matter the lagrangian is
given by
\[
L=L_{\mathrm{ChS}}^{(5)}+\kappa L_{M},
\]
where $L_{\mathrm{ChS}}^{(5)}$ is the five-dimensional Chern--Simons
la\-gran\-gian given by (\ref{1}), $L_{M}=L_{M}(e^{a},h^{a},\omega^{ab})$ is
the matter lagrangian and $\kappa$ is a coupling constant related to the
effective Newton's constant.

The Lagrangian (\ref{1}) shows that standard, five-dimensional General
Relativity emerges as the $l\rightarrow0$ limit of a Chern--Simons theory for
the generalized Poincar\'{e} algebra $B_{5}$. Here $l$ is a length scale, a
coupling constant that characterizes different regimes within the
theory.\textbf{ \ }

The variation of the lagrangian (\ref{1}) w.r.t. the dynamical fields vielbein
$e^{a}$, spin connection $\omega^{ab}$, $h^{a}$ and $k^{ab}$, leads to the
following field equations \cite{gomez1}%
\begin{align}
\varepsilon_{abcde}(2\alpha_{3}R^{ab}e^{c}e^{d}+\alpha_{1}l^{2}R^{ab}R^{cd})
&  =\kappa\frac{\delta L_{M}}{\delta e^{e}}\text{,}\label{ecmova}\\
\alpha_{3}l^{2}\varepsilon_{abcde}R^{ab}R^{cd} &  =\kappa\frac{\delta L_{M}%
}{\delta h^{e}}\text{,}\label{ecmovb}\\
\alpha_{3}l^{2}\varepsilon_{abcde}R^{cd}Dh^{e} &  =0\text{,}\label{ecmov}%
\end{align}
where we have considered that the torsion vanishes $T^{a}=0$ ($\frac{\delta
L_{M}}{\delta\omega^{ab}}=0)$ and $k^{ab}=0$, while the field $h^{a}$ is
associated, in the context of Einstein--Chern-Simons cosmology, with the dark
energy, as shown in Refs. \cite{gomez,gomez1}.

In the case where the equations (\ref{ecmova}-\ref{ecmov}) satisfy the
cosmological principle and the ordinary matter is negligible compared to the
dark energy, we find that the equations (\ref{ecmova} - \ref{ecmov}) take the
form \cite{gomez1}%

\begin{align}
6\left(  \frac{{\dot{a}}^{2}+k}{a^{2}}\right)   &  =\kappa_{5}\alpha\rho
^{(h)},\label{eqz06}\\
3\left[  \frac{\ddot{a}}{a}+\left(  \frac{{\dot{a}}^{2}+k}{a^{2}}\right)
\right]   &  =-\kappa_{5}\alpha p^{(h)},\label{eqz07}\\
{\frac{3l^{2}}{\kappa_{5}}\left(  \frac{{\dot{a}}^{2}+k}{a^{2}}\right)  ^{2}}
&  =\rho^{(h)},\label{eqz08}\\
\frac{3l^{2}}{\kappa_{5}}\frac{\ddot{a}}{a}\left(  \frac{{\dot{a}}^{2}%
+k}{a^{2}}\right)   &  =-p^{(h)},\label{eqz09}\\
\left(  \frac{{\dot{a}}^{2}+k}{a^{2}}\right)  \left[  (h-h(0))\frac{\dot{a}%
}{a}+\dot{h}\right]   &  =0.\label{eqz10}%
\end{align}
The field equations (\ref{eqz06}-\ref{eqz10}) were completely resolved for the
age of Dark Energy in Ref. \cite{gomez1}, where was find that the field
$h^{a}$ has a similar behavior to that of a cosmological constant.

In fact, in Section 3 of Ref. \cite{gomez1} has been found solutions that
describes accelerated expansion for the three possible cosmological models of
the universe. Namely, spherical expansion $\left(  k=1\right)  $, flat
expansion $\left(  k=0\right)  $ and hyperbolic expansion $\left(
k=-1\right)  $. This means that the Einstein--Chern--Simons field equations
have as one of their solutions an universe in accelerated expansion. This
result allow us to conjeture that this solutions are compatible with the era
of Dark Energy and that the energy-momentum tensor for the field $h^{a}$
corresponds to a kind of positive cosmological constant.

\ Introducing (\ref{ecmovb}) in (\ref{ecmova}) we find%
\begin{equation}
\ast\varepsilon_{abcde}(2\alpha_{3}R^{ab}e^{c}e^{d})=\left(  \beta_{1}%
\Upsilon_{e}-\frac{\alpha_{1}}{\alpha_{3}}\beta_{2}\Upsilon_{e}^{(h)}\right)
\text{,} \label{ecmov02}%
\end{equation}
where%
\[
\beta_{1}\Upsilon_{a}=\kappa\ast\left(  \frac{\delta L_{M}}{\delta e^{a}%
}\right)  \text{, \ \ }\beta_{2}\Upsilon_{a}^{(h)}=\kappa\ast\left(
\frac{\delta L_{M}}{\delta h^{a}}\right)  \text{,}%
\]
and $\ast$ is the Hodge star operator.

In the limit of weak gravitational field one assumes that the world metric
tensor $g_{\mu\nu}$ is not very much different from the Minkowski metric
$\eta_{\mu\nu}=\mathrm{diag}(-1,1,...,1)$. In fact, it can be then written in
the form $g_{\mu\nu}=\eta_{\mu\nu}+h_{\mu\nu}\,$, where $h_{\mu\nu}$
represents the small corrections to the flat space-time metric $\eta_{\mu\nu}$
due to the presence of a weak gravitational field. In this approximation
$\left\vert h_{\mu\nu}\right\vert <<1,$ so that terms of order higher than the
first in $h_{\mu\nu}$ can be neglected in the field equations.\ So that,%
\begin{align*}
ds^{2} &  =g_{\mu\nu}dx^{\mu}dx^{\nu}=\eta_{ab}e^{a}e^{b}\\
&  =-(1-h_{00})dt^{2}+(1+h_{11})(dx^{1})^{2}+(1+h_{11})(dx^{2})^{2}%
+(1+h_{11})(dx^{3})^{2}+(1+h_{11})(dx^{4})^{2}\text{.}%
\end{align*}
Introducing an orthonormal basis%
\begin{align*}
e^{0} &  \approx\left(  1+\frac{h_{\text{ \ }0}^{0}}{2}\right)  dt\text{,
\ }e^{1}\approx\left(  1+\frac{h_{\text{ \ }1}^{1}}{2}\right)  dx^{1}\text{,
\ }e^{2}\approx\left(  1+\frac{h_{\text{ \ }2}^{2}}{2}\right)  dx^{2}%
\text{,}\\
e^{3} &  \approx\left(  1+\frac{h_{\text{ \ }3}^{3}}{2}\right)  dx^{3}\text{,
}e^{4}\approx\left(  1+\frac{h_{\text{ \ }4}^{4}}{2}\right)  dx^{4}\text{,}%
\end{align*}
and using the first and second structural equations $T^{a}=de^{a}%
+\omega_{\text{ }b}^{a}e^{b}=0$, \ $R_{\text{ }b}^{a}=d\omega_{\text{ }b}%
^{a}+\omega_{\text{ }c}^{a}\omega_{\text{ }b}^{c}$ we have%
\begin{align*}
R_{00} &  =-\frac{{\nabla}^{2}h_{00}}{2}\text{,}\\
R_{11} &  =\frac{{\partial_{1}}^{2}h_{00}}{2}-\frac{{\partial_{2}}^{2}h_{11}%
}{2}-\frac{{\partial_{3}}^{2}h_{11}}{2}-\frac{{\partial_{4}}^{2}h_{11}}%
{2}\text{,}\\
R_{22} &  =\frac{{\partial_{2}}^{2}h_{00}}{2}-\frac{{\partial_{1}}^{2}h_{22}%
}{2}-\frac{{\partial_{3}}^{2}h_{22}}{2}-\frac{{\partial_{4}}^{2}h_{22}}%
{2}\text{,}\\
R_{33} &  =\frac{{\partial_{3}}^{2}h_{00}}{2}-\frac{{\partial_{1}}^{2}h_{33}%
}{2}-\frac{{\partial_{2}}^{2}h_{33}}{2}+\frac{{\partial_{4}}^{2}h_{33}}%
{2}\text{,}\\
R_{44} &  =\frac{{\partial_{4}}^{2}h_{00}}{2}-\frac{{\partial_{1}}^{2}h_{44}%
}{2}-\frac{{\partial_{2}}^{2}h_{44}}{2}-\frac{{\partial_{3}}^{2}h_{44}}%
{2}\text{.}%
\end{align*}

From (\ref{ecmov02}) we can see%
\begin{equation}
R_{\mu\nu}-\frac{1}{2}g_{\mu\nu}R=-\frac{1}{8\alpha_{3}}\left(  \beta
_{1}\Upsilon_{\mu\nu}-\frac{\alpha_{1}}{\alpha_{3}}\beta_{2}\Upsilon_{\mu\nu
}^{(h)}\right)  \text{.} \label{50a}%
\end{equation}

In the limit of weak gravitational field one assumes that the leading term in
the energy-momentum tensors are $\Upsilon_{00}=\rho$ and $\Upsilon_{00}%
^{(h)}=$ $\rho^{(h)}$ so that%
\begin{equation}
R_{00}=\frac{1}{12\alpha_{3}}\left(  \beta_{1}\rho-\frac{\alpha_{1}}%
{\alpha_{3}}\beta_{2}\rho^{(h)}\right)  \text{.} \label{ecmovfinal}%
\end{equation}

On the another hand the motion of a particle described by the geodesic
equation%
\begin{equation}
\frac{d^{2}x^{\mu}}{ds^{2}}+\Gamma_{\nu\rho}^{\mu}\frac{dx^{\nu}}{ds}%
\frac{dx^{\rho}}{ds}=0\text{,} \label{geo}%
\end{equation}
where $x^{\mu}=\{x^{0},x^{i}\}=(t,x^{i})$. In the nonrelativistic limit eq.
(\ref{geo}) becomes%
\begin{equation}
\frac{d^{2}x^{\mu}}{dt^{2}}=-\Gamma_{00}^{\mu}\left(  \frac{dx^{0}}%
{dt}\right)  ^{2}=-\Gamma_{00}^{\mu}\text{,} \label{geo1}%
\end{equation}
and in the limit of weak gravitational field we can put $g_{\mu\nu}=\eta
_{\mu\nu}+h_{\mu\nu}$, with $\left\vert h_{\mu\nu}\right\vert <<1,$ and we can
neglect terms of order $h^{2}$ and higher. From the definition of the
Christoffel symbols we have $\Gamma_{00}^{i}=-\frac{1}{2}\delta_{\text{ }%
j}^{i}\partial_{j}h_{00}$, where we have assumed that the field is static,
i.e., $\partial_{0}g_{\mu\nu}=0.$ So that the geodetic equation is given by%
\begin{equation}
\frac{d^{2}x^{i}}{dt^{2}}=\frac{1}{2}\delta_{\text{ }j}^{i}\partial_{j}%
h_{00}\text{,} \label{geo3}%
\end{equation}
which coincides with Newton equation of motion $\frac{d^{2}x^{i}}{dt^{2}%
}=-\partial_{i}\phi$ provided that $h_{00}=-2\phi$, therefore $\Gamma_{00}%
^{i}=\delta^{ij}\partial_{j}\phi$. This means that \ the only non-zero
component of the Riemann tensor corresponding to connection $\Gamma_{00}^{i}$
is given by $R_{0j0}^{i}=\delta^{ik}\partial_{k}\partial_{j}\phi$, so from
(\ref{ecmovfinal}) we can conclude that \ the nonrelativistic limit of the
five dimensional Einstein--Chern--Simons gravity is a modified version of the
Poisson equation given by%
\begin{equation}
\nabla^{2}\phi=\frac{2}{3}\left(  k_{1}\rho-\alpha k_{2}\rho^{(h)}\right)
\text{,} \label{riem2'}%
\end{equation}
where $k_{1}=\frac{\beta_{1}}{8\alpha_{3}}$, $k_{2}=\frac{\beta_{2}}%
{24\alpha_{3}}$ and $\alpha=\frac{3\alpha_{1}}{\alpha_{3}}$ \cite{gomez}. If
in (\ref{riem2'}) we choose $\alpha=0$ or $k_{2}=0$ we obtain the Poisson
equation in five dimensions provided that $k_{1}=8\pi G$.

\section{\textbf{Newton--Chern--Simons gravity }}

In Ref. \cite{newton} was shown how the Newton--Cartan formulation of
Newtonian gravity can be obtained from gauging the Bargmann algebra. In Refs.
\cite{gr-chs} was shown that the gauging of $\mathfrak{B}_{5}$ lead to a
five-dimensional Chern--Simons gravity \ which empties into general relativity
in a certain limit. On the other hand, we have seen that the non-relativistic
version of the $\mathfrak{B}_{5}$ algebra is given by the $\mathcal{G}%
\mathfrak{B}_{_{5}}$ algebra and that the procedure of $S$-expansion allows us
to find the $\mathcal{G}\mathfrak{B}_{_{5}}$ algebra from the Newton Hooke
algebra with central extension. In this Section we show that,\ using an
analogous procedure to that used in Ref. \cite{newton}, it is possible to find
a generalization of the Newtonian gravity.

\subsection{\textbf{Gauging the }$\mathcal{G}\mathfrak{B}_{_{5}}$\textbf{
algebra}}

We start with a one-form gauge connection $A$ valued in the $\mathcal{G}%
\mathfrak{B}_{_{5}}$ algebra is given by%
\begin{align}
A  &  =\frac{v}{l}\tau H+\frac{1}{l}e^{i}P_{i}+\frac{v}{l}h^{0}Z_{0}+\frac
{1}{l}h^{i}Z_{i}+\frac{1}{vl}mM+\frac{1}{vl}nN\nonumber\\
&  +\frac{1}{v}\omega^{i}K_{i}+\frac{1}{v}k^{i}Z_{i0}+\frac{1}{2}\omega
^{ij}J_{ij}+\frac{1}{2}k^{ij}Z_{ij}%
\end{align}
where $l$ and $v$ are parameters of dimension length and velocity
respectively. The corresponding two-form curvature is given by%

\begin{align}
F &  =\frac{v}{l}R(H)H+\frac{1}{l}R^{i}(P_{i})P_{i}+\frac{v}{l}R(Z_{0}%
)Z_{0}+\frac{1}{l}R^{i}(Z_{i})Z_{i}+\frac{1}{vl}R(M)M\nonumber\\
&  +\frac{1}{vl}R(N)N+\frac{1}{v}R^{i}\left(  K_{i}\right)  K_{i}+\frac{1}%
{v}R^{i}\left(  Z_{i0}\right)  Z_{i0}+\frac{1}{2}R^{ij}\left(  J_{ij}\right)
J_{ij}+\frac{1}{2}R^{ij}\left(  Z_{ij}\right)  Z_{ij},
\end{align}
\newline where
\begin{align}
R(H) &  =d\tau\text{, }R^{i}(P_{i})=T^{i}-\omega^{i}\tau\text{,}\nonumber\\
R(Z_{0}) &  =dh^{0}\text{, \ }R(M)=dm-\omega^{i}e_{i}\text{,}\nonumber\\
R^{i}(Z_{i}) &  =Dh^{i}-\omega^{i}h^{0}-k^{i}\tau+k_{\,\,j}^{i}e^{j}%
\text{,}\nonumber\\
R(N) &  =dn-\omega^{i}h_{i}-k^{i}e_{i}\text{,}\nonumber\\
R^{i}(Z_{i0}) &  =Dk^{i}+e^{i}\tau+k_{\,\,j}^{i}\omega^{j}\text{, \ }%
R^{i}(K_{i})=D\omega^{i}\text{,}\nonumber\\
R^{ij}(J_{ij}) &  =R^{ij}\text{, \ \ }R^{ij}(Z_{ij})=Dk^{ij}\text{,}%
\end{align}
with $T^{i}=de^{i}+\omega^{ij}e_{j}$, $R^{ij}=d\omega^{ij}+\omega_{\,\,k}%
^{i}\omega^{kj}$.

Since the gauge connection $A$ transforms as
\[
\delta A=d\Lambda+\left[  A,\Lambda\right]  ,
\]
where%
\begin{align*}
\Lambda &  =\frac{v}{l}\zeta^{0}H+\frac{1}{l}\zeta^{i}P_{i}+\frac{v}{l}%
\rho^{0}Z_{0}+\frac{1}{l}\rho^{i}Z_{i}+\frac{1}{vl}\sigma M+\frac{1}{vl}\gamma
N\\
&  +\frac{1}{v}\lambda^{i}K_{i}+\frac{1}{v}\chi^{i}Z_{i0}+\frac{1}{2}%
\lambda^{ij}J_{ij}+\frac{1}{2}\chi^{ij}Z_{ij}\text{,}%
\end{align*}
we find, using the $\mathcal{G}\mathfrak{B}_{_{5}}$ algebra, that the
variations of the gauge fields are given by%
\begin{align}
\delta\tau &  =d\zeta^{0}\text{, \ }\delta e^{i}=D\zeta^{i}-\omega^{i}%
\zeta^{0}-\lambda^{ij}e_{j}+\tau\lambda^{i}\text{,}\nonumber\\
\delta h^{i} &  =D\rho^{i}-\omega^{i}\rho^{0}-\lambda^{ij}h_{j}+h^{0}%
\lambda^{i}+k^{ij}\zeta_{j}-k^{i}\zeta^{0}-\chi^{ij}e_{j}+\tau\chi^{i}%
\text{,}\nonumber\\
\delta m &  =d\sigma-\omega^{i}\zeta_{i}+e^{i}\lambda_{i}\text{, \ }%
\delta\omega^{i}=D\lambda^{i}-\lambda^{ij}\omega_{j}\text{,}\nonumber\\
\delta n &  =d\gamma-k^{i}\zeta_{i}+h^{i}\lambda_{i}-\omega^{i}\rho_{i}%
+e^{i}\chi_{i}\text{,\ }\delta h^{0}=d\rho^{0}\text{,}\nonumber\\
\delta k^{i} &  =D\chi^{i}-\lambda^{ij}k_{j}-\chi^{ij}\omega_{j}+k^{ij}%
\lambda_{j}+e^{i}\zeta^{0}-\zeta^{i}\tau\text{, \ }\delta\omega^{ij}%
=D\lambda^{ij}\text{,}\nonumber\\
\delta k^{ij} &  =D\chi^{ij}+k_{\,\,k}^{i}\lambda^{kj}+k_{\,\,k}^{j}%
\lambda^{ik}\text{,}\label{campos}%
\end{align}
where the derivative $D$ is covariant with respect to the $J$-transformations.

Following Ref. \cite{newton} we impose now several curvature constraints.
These constraints convert the $P$ and $H$ transformations into general
coordinate transformations in space and time. We write the parameter of the
general coordinate transformations $\xi^{\lambda}$ as%
\[
\xi^{\lambda}=e_{i}^{\lambda}\zeta^{i}+\tau^{\lambda}\zeta^{0}.
\]

Here we have used the inverse spatial vielbein $e_{\,\,i}^{\lambda}$ and the
inverse temporal vielbein $\tau^{\lambda}$ defined by \cite{newton}%
\begin{align}
e_{\text{ \ }\mu}^{i}e_{\text{ }\,j}^{\mu}  &  =\delta_{\,\,j}^{i}\text{,
}\,\,\tau^{\mu}\tau_{\mu}=1\text{, \ \ }\tau^{\mu}e_{\text{ \ }\mu}%
^{i}=0\,\nonumber\\
\,\tau_{\mu}e_{\,\ i}^{\mu}  &  =0\text{, \ \ }e_{\text{ \ }\mu}^{i}%
e_{\,\,i}^{\nu}=\delta_{\,\ \mu}^{\nu}-\tau_{\mu}\tau^{\nu}\text{.}%
\end{align}

From (\ref{campos}) we can see that only the gauge fields $e_{\mu}^{\,\,i}$,
$\tau_{\mu}$, $m_{\mu}$, $h_{\mu}^{\,\,i}$, $h_{\mu}^{0}$ and $n_{\mu}$
transform under \ the $P$ and $H$ transformations. These are the fields which
should remain independents, while the remaining fields will be dependent upon
the aforementioned fields. This can be achieved with the following constraints%
\begin{align}
R(H)  &  =d\tau=0\text{, \ }R^{i}(P_{i})=T^{i}-\omega^{i}\tau=0\text{,}\\
R(M)  &  =dm-\omega^{i}e_{i}=0\text{, \ }R(Z_{0})=dh^{0}=0\text{,}\nonumber\\
R^{i}(Z_{i})  &  =Dh^{i}-\omega^{i}h^{0}-k^{i}\tau+k_{j}^{i}e^{j}%
=0\text{,}\nonumber\\
R(N)  &  =dn-\omega^{i}h_{i}-k^{i}e_{i}=0\text{.} \label{constraints}%
\end{align}

An analogous procedure to that used in Ref. \cite{newton} allows us to obtain
the ${k_{\mu}}^{ij}$ and ${k_{\mu}}^{i}$ fields. In fact, using the
constraints (\ref{constraints}) we find%

\begin{equation}
\omega_{\mu}^{\,\,\,\,ij}=(\partial_{\lbrack\mu}e_{\nu]})^{i}e^{\nu
j}-(\partial_{\lbrack\mu}e_{\nu]})^{j}e^{\nu i}+e_{\mu k}(\partial_{\lbrack
\nu}e_{\rho]})^{k}e^{\nu i}e^{\rho j}-\tau_{\mu}e^{\nu\lbrack i}\omega_{\nu
}^{\,\,\,\,j]},
\end{equation}

\begin{equation}
\omega_{\mu}^{\,\,\,\,i}=e^{\nu i}\partial_{\lbrack\mu}m_{\nu]}+e^{\nu i}%
\tau^{\rho}e_{\mu j}{\partial_{\lbrack\nu}e_{\rho]}}^{j}+\tau_{\mu}\tau^{\nu
}e^{\rho i}\partial_{\lbrack\nu}m_{\rho]}+\tau^{\nu}{\partial_{\lbrack\mu
}e_{\nu]}}^{i},
\end{equation}

\begin{align}
k_{\mu}^{\,\,\,\,ij}  &  =(D_{[\mu}h_{\nu]})^{i}e^{\nu j}-(D_{[\mu}h_{\nu
]})^{j}e^{\nu i}+e_{\mu k}(D_{[\nu}h_{\rho]})^{k}e^{\nu i}e^{\rho
j}\nonumber\\
&  -\omega_{\lbrack\mu}^{i}h_{\nu]}^{0}e^{\nu j}+\omega_{\lbrack\mu}^{j}%
h_{\nu]}^{0}e^{\nu i}-e_{\mu k}\omega_{\lbrack\nu}^{k}h_{\nu]}^{0}e^{\nu
i}e^{\rho j}-\tau_{\mu}e^{\nu\lbrack i}k_{\nu}^{\,\,\,\,j]}\text{,}
\label{kij}%
\end{align}%
\begin{align}
k_{\mu}^{\,\,\,\,i}=  &  e^{\nu i}\partial_{\lbrack\mu}n_{\nu]}-e^{\nu
i}{\omega^{k}}_{[\mu}h_{\nu]k}+e^{\nu i}\tau^{\rho}e_{\mu j}{D_{[\nu}h_{\rho
]}}^{j}-e^{\nu i}\tau^{\rho}e_{\mu j}{\omega^{j}}_{[\nu}{h^{0}}_{\rho
]}\nonumber\\
&  +\tau_{\mu}\tau^{\nu}e^{\rho i}\partial_{\lbrack\nu}n_{\rho]}-\tau_{\mu
}\tau^{\nu}e^{\rho i}{\omega^{k}}_{[\nu}h_{\rho k]}+\tau^{\nu}{D_{[\mu}%
h_{\nu]}}^{i}-\tau^{\nu}{\omega^{i}}_{[\mu}{h^{0}}_{\nu]}\text{.} \label{ki0}%
\end{align}

\subsection{\textbf{Newton--Chern--Simons Lagrangian}}

A Chern--Simons lagrangian form $L_{\mathrm{ChS}}(A,0)\equiv Q_{2n+1}(A,0)$ is
a diferential form defined for a connection, whose exterior derivative yields
a Chern class. Although the Chern classes are gauge invariant, the
Chern--Simons forms are not; under gauge transformations they change by a
closed form.\ A transgression form $Q_{2n+1}(A_{1},A_{2})$ on the other hand,
is an invariant differential form whose exterior derivative is the difference
of two Chern classes. \ It generalizes the Chern--Simons form with the
additional advantage that it is gauge invariant. \ 

To obtain the lagrangian for $5$-dimensional Chern--Simons gravity we use
subspaces separation method introduced in Ref. \cite{salg05} and write
$L_{\mathrm{ChS}}$ in terms of a transgression form, a Chern--Simons form and
a total exact form
\[
Q_{5}(A_{1},0)=Q_{5}(A_{1},A_{2})+Q_{5}(A_{2},0)+dQ_{4}(A_{1},A_{2},0),
\]
where,%

\begin{equation}
Q_{5}(A_{1},A_{2})=3\int_{0}^{1}dt\left\langle \theta F_{t}^{2}\right\rangle
\end{equation}
with $\theta=A_{1}-A_{2},$ $A_{t}=A_{2}+t\theta,$ $A_{1}=A,$ $A_{2}%
=\omega=\frac{1}{2}\omega^{ij}J_{ij}$, $F_{t}=dA_{t}+A_{t}A_{t}$ and%

\begin{equation}
Q_{5}(A_{2},0)=3\int_{0}^{1}dt\left\langle A_{2}F_{t}^{2}\right\rangle
\end{equation}
where now $A_{t}=tA_{2}=t\omega.$

\bigskip So that if we don't consider boundary terms the Chern-Simons
lagrangian is given by:%
\begin{align}
L_{\mathrm{ChS},\mathcal{G}\mathfrak{B}_{_{5}}}  &  =\alpha_{1}\epsilon
_{ijkl}\left(  -2R^{ij}T^{k}\omega^{l}-\frac{4}{3}R^{ij}\omega^{k}\omega
^{l}\tau+2R^{ij}D\omega^{k}e^{l}-R^{ij}R^{kl}m\right)  +\alpha_{3}%
\epsilon_{ijkl}\left(  \frac{4}{3}\nu^{2}R^{ij}e^{k}e^{l}\tau\right.
\nonumber\\
&  \,\,\,\,\left.  -2R^{ij}Dh^{k}\omega^{l}-\frac{4}{3}R^{ij}k^{k}\omega
^{l}\tau-\frac{4}{3}R^{ij}\omega^{k}\omega^{l}\hat{\tau}+2R^{ij}D\omega
^{k}h^{l}-\frac{4}{3}Dk^{ij}T^{k}\omega^{l}\right. \nonumber\\
&  \left.  -Dk^{ij}\omega^{k}\omega^{l}\tau-R^{ij}k^{kl}dm-\frac{2}{3}%
R^{ij}k^{kl}e^{m}\omega_{m}-\frac{2}{3}R^{ij}\omega_{\,m}^{k}k^{ml}m-\frac
{4}{3}k^{ij}T^{k}D\omega^{l}\right. \nonumber\\
&  \,\,\,\,\left.  -k^{ij}D\omega^{k}\omega^{l}\tau-2R^{ij}T^{k}k^{l}-\frac
{4}{3}R^{ij}\omega^{k}k^{l}\tau+\frac{2}{3}R^{ij}k^{km}\omega_{m}e^{l}%
+\frac{2}{3}\omega_{\,m}^{i}k^{jm}D\omega^{k}e^{l}\right. \nonumber\\
&  \,\,\,\,\left.  -R^{ij}R^{kl}n-2R^{ij}\omega^{km}k_{m}e^{l}\right)
\text{.} \label{lagr}%
\end{align}

The lagrangian variation of (\ref{lagr}) leads to the following equations of
motion:%
\begin{align}
\epsilon_{ijkl}(-\frac{4}{3}\alpha_{1}R^{ij}\omega^{k}\omega^{l}+\frac{4}%
{3}\nu^{2}\alpha_{3}R^{ij}e^{k}e^{l})  &  =\kappa\frac{\delta L_{M}}%
{\delta\tau}\text{,}\label{mov1}\\
\frac{4}{3}\alpha_{3}\epsilon_{ijkl0}R^{ij}\omega^{k}\omega^{l}  &
=-\kappa\frac{\delta L_{M}}{\delta\hat{\tau}}\text{,}\label{mov2}\\
4\epsilon_{ijkl}\left(  \alpha_{1}R^{ij}D\omega^{k}-\frac{2}{3}\nu^{2}%
\alpha_{3}R^{ij}e^{k}\tau\right)   &  =\kappa\frac{\delta L_{M}}{\delta e^{l}%
}\text{,}\label{mov3}\\
2\alpha_{3}\epsilon_{ijkl}R^{ij}D\omega^{k}  &  =\kappa\frac{\delta L_{M}%
}{\delta h^{l}}\text{,}\label{mov4}\\
\alpha_{1}\epsilon_{ijkl}R^{ij}R^{kl}  &  =-\kappa\frac{\delta L_{M}}{\delta
m}\text{,}\label{mov5}\\
\alpha_{3}\epsilon_{ijkl}R^{ij}R^{kl}  &  =-\kappa\frac{\delta L_{M}}{\delta
n}\text{,}\label{mov6}\\
4\epsilon_{ijkl}\left(  \frac{2\alpha_{1}}{3}R^{ij}\omega^{k}\tau-\alpha
_{1}R^{ij}T^{k}+\frac{2\alpha_{3}}{3}R^{ij}\omega^{k}\hat{\tau}-\alpha
_{3}R^{ij}Dh^{k}\right)   &  =\kappa\frac{\delta L_{M}}{\delta\omega^{l}%
}\text{,} \label{mov7}%
\end{align}%
\begin{gather}
\epsilon_{ijkl}\left(  -2\alpha_{1}R^{km}e_{m}\omega^{l}-4\alpha_{1}%
T^{k}D\omega^{l}-\frac{4\alpha_{1}}{3}\omega^{k}\omega^{l}d\tau-\frac
{8\alpha_{1}}{3}D\omega^{k}\omega^{l}\tau+2\alpha_{1}R^{km}\omega_{m}%
e^{l}\right. \nonumber\\
-2\alpha_{1}R^{kl}dm+\frac{8}{3}\nu^{2}\alpha_{3}T^{k}e^{l}\tau+\frac
{4\alpha_{3}}{3}e^{k}e^{l}d\tau-2\alpha_{3}R^{km}h_{m}\omega^{l}-4\alpha
_{3}Dh^{k}D\omega^{l}\nonumber\\
\left.  -\frac{4\alpha_{3}}{3}\omega^{k}\omega^{l}d\hat{\tau}-\frac
{8\alpha_{3}}{3}D\omega^{k}\omega^{l}\hat{\tau}+2\alpha_{3}R^{km}\omega
_{m}h^{l}-\alpha_{3}R^{kl}dn\right)  =\kappa\frac{\delta L_{M}}{\delta
\omega^{ij}}\text{,} \label{mov8}%
\end{gather}
where, we have considered, in analogy with the Section IV, $k^{ij}=k^{i}=0.$
The first four equations corresponding to the non-relativistic version of the
Einstein equations. The equations (\ref{mov5}) and (\ref{mov6}) are second
order curvatures then in the limit of weak gravitational field $\frac{\delta
L_{M}}{\delta m}=\frac{\delta L_{M}}{\delta n}=0$. The equations (\ref{mov7})
and (\ref{mov8}) corresponding to the non-relativistic version of the torsion equation.

In analogy with the Section IV, the first two lead us to%
\begin{equation}
\ast\epsilon_{ijkl}R^{ij}e^{k}e^{l}=\frac{3}{4\nu^{2}}\left(  \frac{\beta_{1}%
}{\alpha_{3}}\Upsilon_{0}-\frac{\alpha_{1}\beta_{2}}{{\alpha_{3}}^{2}}%
\Upsilon_{0}^{(h)}\right)  =\Omega_{0}\text{,}\label{poisson3}%
\end{equation}
where we found that $4R_{00}=\Omega_{00}$ with $R_{00}=\nabla^{2}\phi$.
Finally from (\ref{poisson3}) we obtain%
\begin{equation}
\nabla^{2}\phi=\frac{3}{2\nu^{2}}(k_{1}\rho-\alpha k_{2}\rho^{(h)}%
)\text{,}\label{riem2''}%
\end{equation}
where the constants $k_{1}=\frac{\beta_{1}}{8\alpha_{3}}=8\pi G$, $k_{2}%
=\frac{\beta_{2}}{24\alpha_{3}}$ and $\alpha=\frac{3\alpha_{1}}{\alpha_{3}}$.
This result coincides with the equation (\ref{riem2'}) if $\nu=\frac{3}{2}$.
\ This results shows that the non-relativistic limit of
Einstein--Chern--Simons gravity, invariant under the $\mathfrak{B}_{5}$
algebra coincides with Newton --Chern--Simons gravity invariant under the
algebra $\mathcal{G}\mathfrak{B}_{_{5}}$.

\section{\textbf{Comments }}

In the present work we have shown that:\ $\mathbf{(i)}$ it is possible to
obtain the non-relativistic versions of both generalized Poincar\'{e} algebras
and generalized $AdS$-Lorentz algebras. These non-relativistic algebras are
called generalized Galilean type I and type II algebras and denoted by
$\mathcal{G}\mathfrak{B}_{_{n}}$ and $\mathcal{G}\mathfrak{L}_{_{n}}$
respectively.\ $\mathbf{(ii)}$ The procedure of $S$-expansion allows us to
find the $\mathcal{G}\mathfrak{B}_{_{5}}$ algebra from the Newton--Hooke
algebra with central extension. $\mathbf{(iii)}$ Using an analogous procedure
to that used in Ref. \cite{newton}, it is possible to find\ the
non-relativistic limit of the five dimensional Einstein--Chern--Simons gravity
which lead us to a modified version of the Poisson equation.

It is interesting to note that the $\mathfrak{B}_{5}$ algebra is a
generalization of the Poincar\'{e} algebra which includes the extra generators
$Z_{ab}$ and $Z_{a}$. This algebra leads to a Chern--Simons lagrangian which
coincides with the Einstein--Hilbert lagrangian in a certain limit, even if
the new gauge field vanishes and therefore leads to newtonian gravity in the
non relativistic limit. The generators $Z_{i0},Z_{ij}$, are the space-time
components of the $Z_{ab}=\left(  Z_{i0},Z_{ij}\right)  $ relativistic
generators, whose gauge field $k^{ab}=\left(  k^{i},k^{ij}\right)  $ we fix to
$k^{ab}=0$ in the field equations.

On the other hand the gauge field $h^{a}=\left(  h^{0},h^{i}\right)  $
associated to the generators $Z_{a}$ generates modifications in the Einstein
equations which can be interpretated, in the cosmological contex, as an effect
due to the dark energy \cite{gomez,gomez1}. \ This modifications leads, in the
non-relativistic limit, to a modification in the Poisson equation shown in
(\ref{riem2''}), which could be compatible with the Dark Matter. This would
allow us conjecture that Dark Matter could be interpreted as the
non-relativistic limit of Dark Energy.

The modified form of Poisson equation (\ref{riem2''}) suggests a possible
connection with\ the so called MOND approach to gravity interactions. In fact
the first complete theory of MOND was constructed by Milgrom and Bekenstein in
Ref. \cite{Milg}. This theory is based on the lagrangian%
\begin{equation}
\mathcal{L}=-\frac{a_{0}^{2}}{8\pi G}f\left(  \frac{\left\vert \vec{\nabla
}\phi\right\vert }{a_{0}^{2}}\right)  -\rho\phi
\end{equation}
where $\phi$ is the gravitational potencial (meaning that for a test particle
$\vec{a}=-\vec{\nabla}\phi$), and $\rho$ denotes the matter mass density. The
corresponding equation for $\phi$ is given by%
\begin{equation}
\vec{\nabla}\cdot\left[  \mu\left(  \frac{\left\vert \vec{\nabla}%
\phi\right\vert }{a_{0}}\vec{\nabla}\phi\right)  \right]  =4\pi G\rho
\end{equation}
where $\mu(\sqrt{y})=df(y)/dy$, which can be written as\textbf{ }%
\begin{equation}
\mu\nabla^{2}\phi=4\pi G\rho-\vec{\nabla}\mu\cdot\vec{\nabla}\phi.
\end{equation}

\textbf{\ }

Comparing this last equation with equation (\ref{riem2''}), we can see that in
some particular cases the MOND approach to gravity could coincide with the
modified Poisson equation (\ref{riem2''})

\begin{acknowledgement}
\textit{This work was supported in part by FONDECYT Grants }$N_{o}$\textit{
1130653 and }$N_{o}$ 1150719 from the Government of Chile.\textit{\ Two of the
authors (}GR, SS\textit{) were supported by grants from Comision Nacional de
Investigaci\'{o}n Cient\'{\i}fica y Tecnol\'{o}gica }CONICYT\textit{ }N$_{o}%
$\textit{ 21140971 and }N$_{o}$ \textit{21140490 respectively and from
Universidad de Concepci\'{o}n, Chile. }NG\textit{ was supported by a grant
from }CONICYT-DAAD\textit{ }N$_{o}$\textit{ A1472364.}
\end{acknowledgement}

\end{document}